\documentclass[traditabstract]{aa} 
\usepackage{graphicx}

\bibpunct{(}{)}{;}{a}{}{,} 

\begin{document}

\title{Wave dynamics in a sunspot umbra}
\author{R. Sych\inst{1,2} \and V.~M.~Nakariakov\inst{3,4,5}\thanks{Corresponding~author: V.~M.~Nakariakov, V.Nakariakov@warwick.ac.uk}}

\authorrunning{Sych \& Nakariakov}
\titlerunning{Wave dynamics in a sunspot umbra}

\institute{Key Laboratory of Solar Activity, National Astronomical Observatories, Chinese Academy of Sciences, A20 Datun Road, Chaoyang District, Beijing, China\\
           \email{sych@iszf.irk.ru}
            \and
           Institute of Solar-Terrestrial Physics, Irkutsk, Lermontov St., 126a, 664033, Russia     
           \and
Centre for Fusion, Space and Astrophysics, Department of Physics, University of Warwick, CV4 7AL, UK\\
\email{V.Nakariakov@warwick.ac.uk}\label{1}
\and
School of Space Research, Kyung Hee University, Yongin, 446-701, Gyeonggi, Korea\label{2}
\and
Central Astronomical Observatory at Pulkovo of the Russian Academy of Sciences, St Petersburg 196140, Russia\label{3}
}

\date{Received \today /Accepted dd mm yyyy}

\abstract
{Sunspot oscillations are one of the most frequently studied wave phenomena in the solar atmosphere. Understanding the basic physical processes responsible for sunspot oscillations requires detailed information about their fine structure.}  
{We aim to reveal the relationship between the fine horizontal and vertical structure, time evolution, and the fine spectral structure of oscillations in a sunspot umbra. }
{The high spatial and time resolution data obtained with SDO/AIA for the sunspot in active region NOAA 11131 on 08 December 2010 were analysed with the time-distance plot technique and the pixelised wavelet filtering method. Different levels of the sunspot atmosphere were studied from the temperature minimum to the corona.}
{Oscillations in the 3 min band dominate in the umbra. The integrated spectrum of umbral oscillations contains distinct narrowband peaks at 1.9~min, 2.3~min, and 2.8~min. The power significantly varies in time, forming distinct 12--20~min oscillation trains. The oscillation power distribution over the sunspot in the horizontal plane reveals that the enhancements of the oscillation amplitude, or wave fronts, have a distinct structure consisting of an evolving two-armed spiral and a stationary circular patch at the spiral origin, situated near the umbra centre. This structure is seen from the temperature minimum at 1700\AA\ to the 1.6~MK corona at 193\AA. In time, the spiral rotates anti-clockwise. The wave front spirality  is most pronounced during the maximum amplitude phases of the oscillations, and in the bandpasses where umbral oscillations have the highest power, 304\AA\ and 171\AA. In the low-amplitude phases the spiral breaks into arc-shaped patches. The 2D cross-correlation function shows that the oscillations at higher atmospheric levels occur later than at lower layers. The phase speed is estimated to be about 100~km/s. The fine spectral analysis shows that the central patch corresponds to the high-frequency oscillations, while the spiral arms highlight the lower-frequency oscillations in the 3-min band. 
}
{The vertical and horizontal radial structure of the oscillations is consistent with the model that interprets umbral oscillations as slow magnetoacoustic waves filtered by the atmospheric temperature non-uniformity in the presence of  the magnetic field inclination from the vertical. The mechanism for the polar-angle structure of the oscillations, in particular the spirality of the wave fronts, needs to be revealed.}


\maketitle

\section{Introduction}
\label{intro}

Sunspot oscillations have been subject to intensive studies for several decades
\citep[see, e.g.][for comprehensive reviews]{2000SoPh..192..373B,2003A&ARv..11..153S,2006RSPTA.364..313B}.
The interest in sunspot oscillations is, first of all, connected with the possibility of sunspot seismology - remote probing of the sunspot structure and physical conditions by the oscillations \citep[see, e.g.][]{2008SoPh..251..501Z}.
In particular, very recently the spatial structure of 3--10 min sunspot oscillations was used in studying the sunspot magnetospheric geometry  \citep{2014A&A...561A..19Y}. Long-period sunspot oscillations may indicate 
the validity of the shallow sunspot model in the sub-photospheric layers \citep{2008SoPh..248..395S, 2013PASJ...65S..13B}. 
Sunspot oscillations are also known to be the source of upward propagating waves in the corona \citep[see, e.g.][]{2012RSPTA.370.3193D, 2012ApJ...757..160J, 2010ApJ...724L.194V}. It makes them directly relevant for the
seismology of the corona, because they provide important information about the input wave spatial and time spectrum. Moreover, the recently established relationship between the processes in sunspots and energy releases in the corona \citep{2009A&A...505..791S} opens up a promising possibility for probing the magnetic connectivity in the solar atmosphere by the leakage of sunspot oscillations to the corona. Effective use of sunspot oscillations for sunspot and coronal seismology requires detailed knowledge of their spatial and spectral structure and time evolution at different atmospheric levels, and the advanced theory.

Theoretical modelling of 3-min oscillations has been a challenge for several decades, and so far, there is no universally accepted theory of this phenomenon. It is commonly believed that 3-min oscillations are caused by slow magnetoacoustic waves \citep{2008SoPh..251..501Z}. In the low-$\beta$ plasma of the umbra, slow waves are practically field-aligned compressive motions of the plasma. But it is debated whether the waves are standing, resonating between two reflective layers \citep[e.g.][]{1997RSPSA.453..943B}, or result from the interaction of upwardly propagating slow waves with the stratified plasma. In the latter model, known as the filter theory, the sunspot chromosphere is considered as a Fabry--Perot filter for the slow waves confined to the strong magnetic field \citep{1981SvAL....7...25Z,1983SoPh...82..369Z}. In this theory, 3-min oscillations are excited by irregular motions outside and below the sunspot, and only waves with specific frequencies are transmitted upwards through the sunspot atmosphere, producing discrete spectral peaks.
Recent numerical modelling \citep{2011ApJ...728...84B} reveals that different profiles of the plasma temperature and density lead to different frequencies and different efficiency of the leakage into the corona. Thus, horizontal non-uniformity of the sunspot atmosphere naturally results in fine spectral and power structuring of 3-min oscillations across the umbra. Moreover, the structuring can be different at different heights. 
There is a clear evidence of significant changes of the 3-min oscillation spectrum across the umbra \citep[see][for discussion and references therein]{2008SoPh..251..501Z}.

Despite the continuously growing empirical knowledge and the increasing complexity of theoretical models, there remain a number of open questions connected with the basic physical mechanisms operating in sunspots. There is a need for better understanding of the fine spatial structure of the oscillations in both the vertical and horizontal directions for different sub-bands of the 3-min spectral band and its time evolution. 
Modern observational facilities, such as the Atmospheric Imaging Assembly on the Solar Dynamics Observatory, SDO/AIA \citep{2012SoPh..275...17L}, open up the possibility for the detailed multi-wavelength investigation of sunspot oscillations with high time and spatial resolution for the time sufficiently long to reveal fine spectral features.
This approach enables the simultaneous detailed study of processes at different heights in the sunspot atmosphere.
It has already been established that the period corresponding to the highest power of sunspot oscillations decreases with the distance from the umbra centre in both intensity oscillations \citep[e.g.][]{2008SoPh..248..395S, 2012ApJ...756...35R, 2014A&A...561A..19Y} and Doppler shift oscillations \citep{2013SoPh..288...73M}, and in their combination
\citep{2013A&A...554A.146K}.
It was also found that the instant oscillation period varies with the amplitude  \citep{2012A&A...539A..23S}. 
High-precision spectral measurements revealed the effect of  3-min oscillation height inversion: in the umbral
3-min oscillations are more suppressed than in the neighbouring regions, whereas the chromospheric oscillations are enhanced \citep{2008AstL...34..133K, 2011A&A...525A..41K}. The propagation channel guiding 3-min oscillations from the photospheric levels to the corona was traced in \cite{2012ApJ...757..160J} and \cite{2013ApJ...762...42S}.

The aim of this paper is to study spatiotemporal and spectral dynamics of wave fronts of 3-min oscillations in
a sunspot umbra, with the main attention on the fine details. The analysed data set is discussed in Section~\ref{observ}, the analysis is presented in Section~\ref{analys}, {the results are summarised in
Section~\ref{res}
and discussed in Section~\ref{disc}}.

\section{Observations}
\label{observ}

We analysed oscillations of the EUV emission generated in a symmetric sunspot that was situated in active region NOAA 11131 during a two-hour interval (02:00-04:00~UT) on 08 December, 2010. The sunspot was in the Northern hemisphere, near the central meridian, {at N31, E01}. This active region existed from 01 December to 24 December, 2010, and had low flaring activity. This active region has been subject to intensive studies, which revealed a spatial localisation of narrowband oscillations in the 2--15~min band \citep{2012ApJ...746..119R,2014A&A...561A..19Y}, established the relationship between the drifts of instant frequency and the power of 3-min oscillations \citep{2012A&A...539A..23S}, and
determined the magnetic field geometry in the sunspot magnetosphere by the new technique based on the acoustic cut-off frequency 
\citep{2014A&A...561A..19Y}. 

In the present study, we used the UV and EUV intensity data cubes obtained with SDO/AIA  in the bandpasses 1700\AA\ (the temperature minimum); 1600\AA\ - the lower chromosphere, \ion{C}{iv} and continuum; 
304\AA\ - the transition region, \ion{He}{ii}; 171\AA\ - the corona, \ion{Fe}{ix}; 193\AA\ - the corona, 
\ion{Fe}{xii}; 211\AA\ - coronal hot active regions, \ion{Fe}{xiv}; and 335\AA\ - coronal hot active regions,
 \ion{Fe}{xvi}. The time resolution was 24 s for 1700\AA\ and 1600\AA, and 12 s for other channels. 
 The pixel size was 0.6\arcsec. The duration of the analysed signal was two hours, which allowed for detecting the oscillations with the periods from 0.5 min to 40 min. 
The region of interest (RoI) was a square of 72\arcsec$\times$72\arcsec centred on the sunspot centre. The RoI included both the umbra, of a size of about 24\arcsec, and the penumbra,  of about 48\arcsec. 
The analysed sunspot had an almost circular shape with a well-developed umbra and penumbra. 
The data cubes were constructed with the use of the informational resource Heliophysics Coverage Registry\footnote{http://www.lmsal.com/get\_aia\_data/} that provides Level-1 calibrated and derotated images. 

\section{Analysis}
\label{analys}

\subsection{1D analysis}

\begin{figure}
   \centering
  \includegraphics[width=9.0 cm]{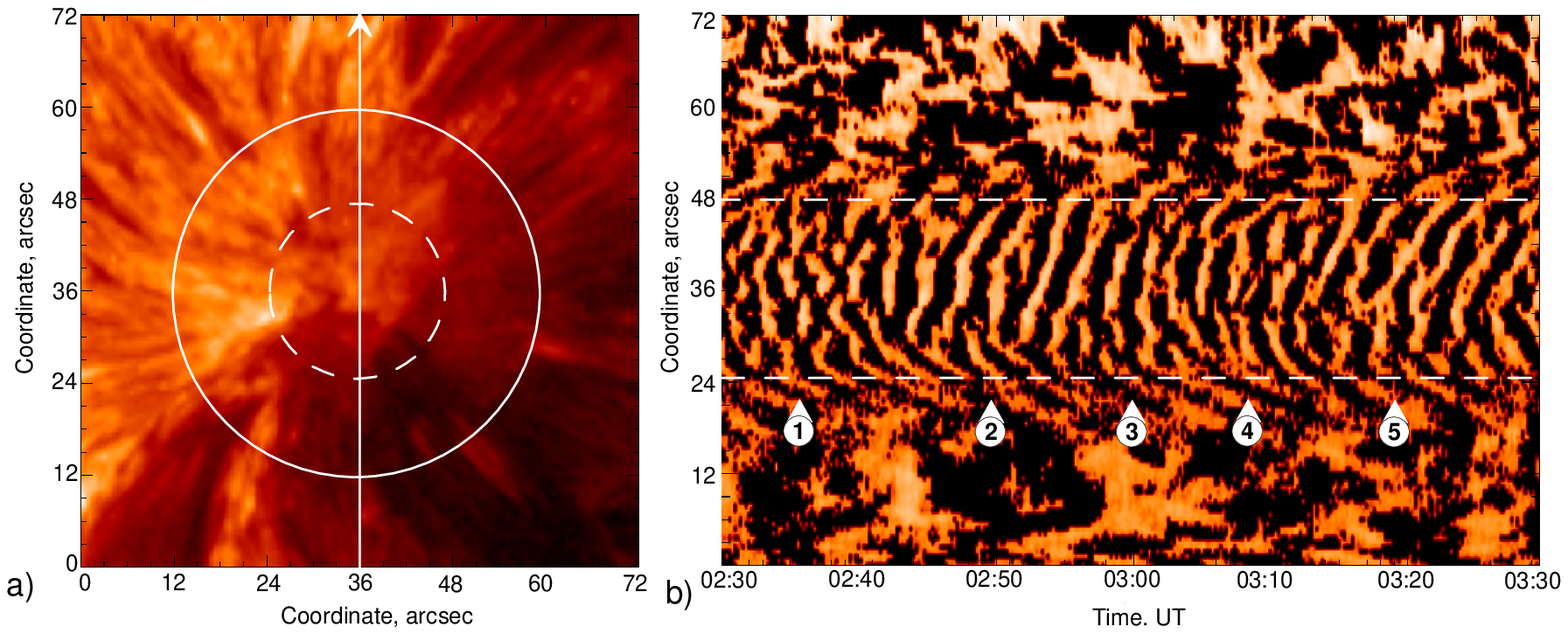}
  \caption{(a) SDO/AIA 304\AA\ image of sunspot NOAA 11131 on 08 December 2010 at 02:00~UT. The dashed line shows the umbra's outer boundary, the solid line shows the outer boundary of the penumbra. The vertical line shows the slit of the time--distance plot. (b) The time--distance plot constructed for the time interval 02:30--03:30~UT in the south--north direction through the sunspot centre. The intensity is given in the logarithmic scale. The horizontal dashed lines show the umbral boundaries. The white triangles labelled with digits show the instants of time when the 3-min oscillation power {reaches the local maxima}.
  }
\label{fig1}
\end{figure}

Preliminary information about oscillatory processes in the sunspot was obtained by constructing a time-distance map of the oscillations observed at 304\AA. Consider a 1D slit taken in the south-north direction (see Fig.~\ref{fig1}a) and passing through the sunspot centre. From the signals of each spatial point, the global trend, determined as a best-fitting sixth-order polynomial, was subtracted.
Then, the spatial non-uniformity of the brightness over the sunspot, for example the decrease in the brightness in the umbra, was removed by subtracting the average values. Fig.~\ref{fig1}b shows the time-distance map constructed by the obtained intensity variations in the logarithmic scale. For the visualisation, we added the lowest value and a unity to the signal of each pixel, and then took a logarithm of the signal. This operation allowed us to avoid taking a logarithm of negative values or a zero.
The time-distance map reveals the oscillatory processes that are non-uniformly distributed over the sunspot. 

In the umbra we see quasi-monochromatic 3-min oscillations with the wave fronts that have a distinct horse-shoe (also known as chevron-like shape, see, e.g. \cite{2006SoPh..238..231K}): the angle of the inclination of the fronts increases towards the umbra boundary. The increase in the inclination angle corresponds to the decrease in the absolute value of the apparent phase speed of the waves. On the either sides of the umbra the wave front inclination has different signs, indicating the change in the sign of the apparent phase velocity. The 3-min oscillations are spatially constrained within the umbra. In the penumbra, 5-min oscillations dominate near the umbra-penumbra boundary. At a larger distance from the umbra-penumbra boundary, there are 15--20-min oscillations. 

\begin{figure}
   \centering
  \includegraphics[width=9.0 cm]{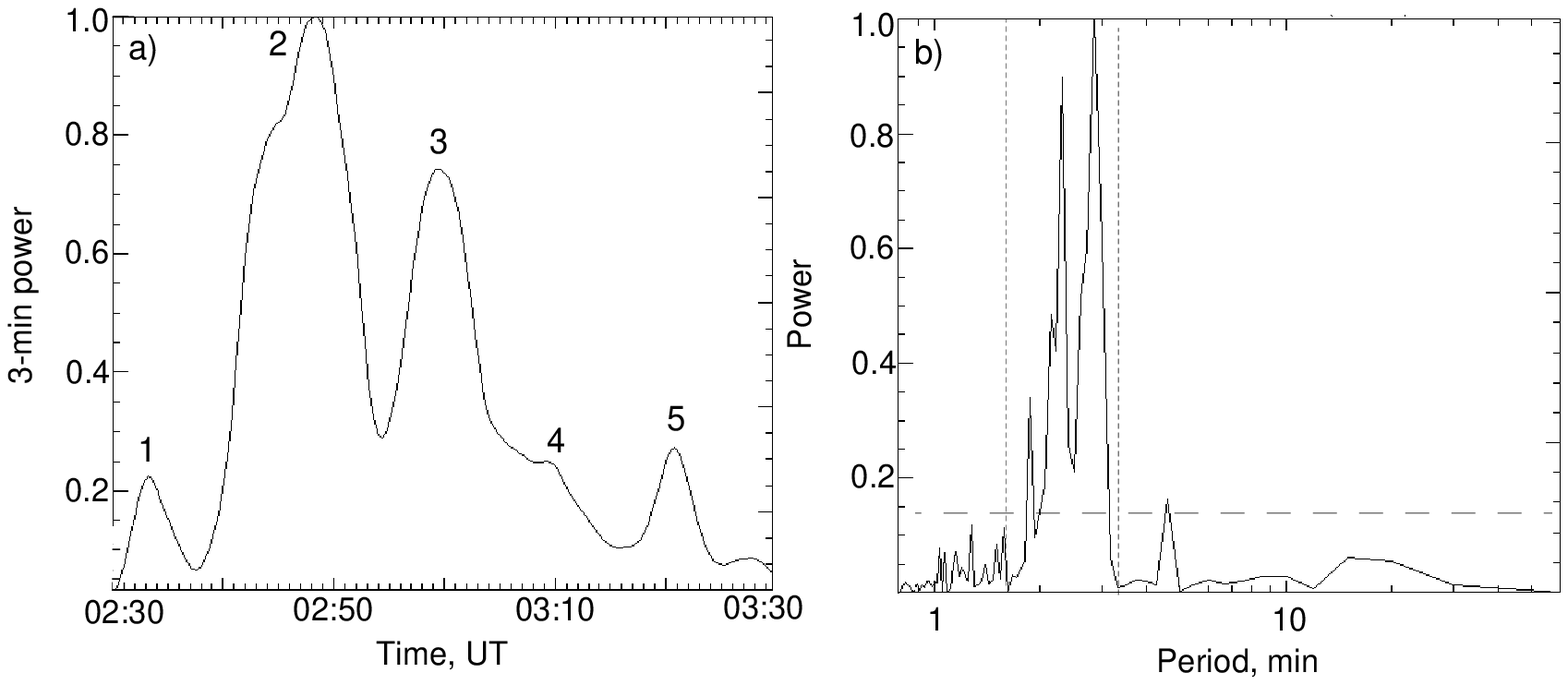}
  \caption{(a) Time variation of 3-min oscillation power {integrated in the range of periods from 1.5 min to 3.5 min in the sunspot umbra. The signal is} measured in the 304\AA\ emission intensity. The digits label the time intervals of the enhanced power {(see also Fig.~\ref{fig1}b)}. (b) {Fourier power spectrum of oscillations of the 304\AA\ emission intensity integrated over} the umbra. {The 3-min power and spectral power density are given in arbitrary units normalised to the highest values.}
  \label{fig2}
} \end{figure}

The sequence of the 3-min wave fronts is not even. There are time intervals (e.g. 02:40-03:00~UT) when the fronts are parallel to each other. On the other hand, there are time intervals (e.g. 02:30-02:40~UT and 03:05-03:15~UT) when the fronts experience breaks and dislocations, and are not parallel to each other. In the time intervals of different spatial coherency {the 3-min oscillation power change as well. This effect is illustrated by the time variation of the 3-min oscillation power shown in Fig.~\ref{fig2}a. {This was determined from the wavelet  power spectrum.
The 304~\AA\ emission intensity signal 
was obtained from the wavelet power spectra of the emission intensity variation at the pixels situated on the slit positioned across the umbra, shown in Fig.~\ref{fig1}a. The wavelet power spectra of the pixels along the slit were summed. In the wavelet power spectrum of the spatially integrated signal, we integrated the spectral power in the 1.5--3.5~min bandpass, and obtained the time signal of interest.}
It is evident that 3-min oscillations form wave trains, or, in other words, their amplitude is modulated with a period of about 12--15 min. The comparison of the time-distance map (Fig.~\ref{fig1}b) and the power curve (Fig.~\ref{fig2}a) reveals certain correlations: in the time intervals of the maximum oscillation power the wave fronts behave more regularly than in the time intervals of the minimum power.

Fig.~\ref{fig2}b shows the power spectrum of the EUV intensity variation integrated over the area within about 5\arcsec\ from the centre of the umbra. The main part of the power is situated in the 3-min band, from 1.5 min to 3.5 min. {In the following analysis and discussion, we refer to this period range from 1.5 min to 3.5 min as a \lq\lq 3-min band\rq\rq}. Inside this band, there are distinct peaks at 1.9 min, 2.3 min and 2.9 min. In addition, there are spectral peaks at 4.7~min and 14.8~min. The 4.7-min peak can be associated with the 5-min oscillations that are suppressed in the umbra, but are enhanced at the umbra-penumbra boundary. {For reference, the 99\% confidence level is shown.}

{In Fig.~\ref{fig2}b the 15-min spectral is lower than 99\% confidence level. But we may consider it as a true periodic signal, as it has already been pointed out in earlier studies \citep[e.g.][]{2010A&A...513A..27C,2012A&A...539A..23S}. The nature of the 15-min peak is not known.} It is interesting that this period coincides with the width of the envelope of the 3-min oscillation peak. There may also be the question whether the peak might be connected with the applied detrending of the signal by subtracting a sixth-order polynomial, which can have up to five extreme points. In one oscillation cycle, there are one maximum and one minimum extreme points. Hence, a sixth-order polynomial can have no more than three oscillation cycles. As the duration of the analysed sample is 120 min, the artificial periodicity introduced by the careless subtraction of a sixth-order polynomial trend would have periods of 40 min or longer. Thus, we disregard any association of the appearance of the 15-min spectral peak with the detrending artefacts.}

\subsection{2D analysis}
\label{2dan}

An analysis of the 2D structure of the oscillations may bring additional important information. We performed it with the pixelised wavelet filtering (PWF) method \citep{2008SoPh..248..395S, 2010SoPh..266..349S}, which allows us to determine the spatiotemporal structure of the spectrum, power, and phase of the oscillations in the sunspot. Moreover, the narrowband filtering of the signal, performed by this technique, improves the signal-to-noise ratio in the spectral band of interest.

We applied the PWF technique to the 304\AA\ data cube, obtained the 4D data cube (two spatial coordinates, time, and frequency), and reduced it to a narrowband 3D data cube (two spatial coordinates and time) for the 3-min band {(the signals with the periods in the range from 1.5 min to 3.5 min)}.
The spatial RoI was reduced with respect to the region shown in Fig.~\ref{fig1}a to a $36''\times 36''$ box centred at the sunspot centre, to speed up the calculations. 
 
\begin{figure}[htpb]
   \centering
  \includegraphics[width=9.0 cm]{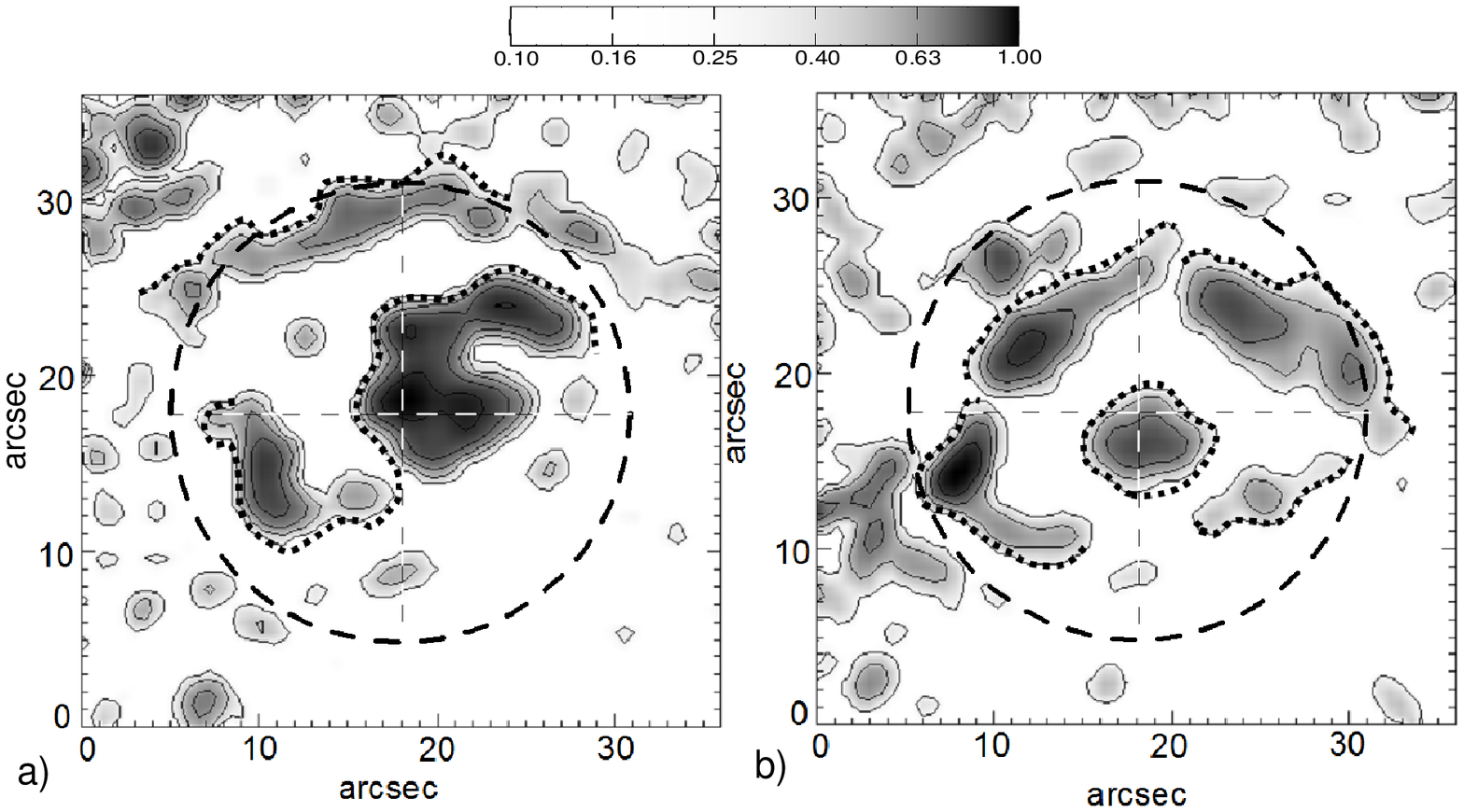}
  \caption{Snapshots of spatial distribution of 3-min oscillation power measured in 304\AA\ in the umbra of sunspot NOAA 11131, {(a) at the time instant of the maximum 3-min oscillation power, 02:49~UT, and (b) at the minimum power, 02:35~UT}, showing the shape of wave fronts. The dashed curve indicates the umbral boundary. The signal is shown in the logarithmic scale. The black dotted curves highlight the apparent wave fronts. {The bar shows the normalised logarithmic power, black indicates the high power, and white the lower power.}}
     \label{fig3}
  \end{figure}

Snapshots of the 3-min narrowband 3D data cube 
taken at different instants of time reveal that the instant shapes of the wave fronts change in time. The most prevailing shape is a two-armed spiral whose origin coincides with the sunspot centre, rotating anti-clockwise. This process is accompanied with the arm broadening, departure of the arms from the central part, and formation of arc-shaped fronts moving towards the umbra boundary, and rapid decay near the umbra boundary. {Fig.~\ref{fig3}a} shows a snapshot of the wave fronts at 02:49~UT, when the oscillation power was highest. The instant narrowband amplitude of the 3-min oscillations is shown in the logarithmic scale, to allow for the comparison of bright and faint details in the image. This amplitude is largest at the centre of the sunspot and decreases rapidly to the umbra boundary. {The black dotted  curves highlight the wave fronts}. Because of this the fronts are not clearly visible in the linear scale and we visualise them in the logarithmic scale. In this approach, the signals situated outside the umbra-penumbra boundary, where the signals are very faint, are meaningless. 

In addition to the spiral shape, the fronts have a patchy circular shape in some time intervals that surrounds the source localised at the umbra centre (see Fig.~\ref{fig3}b). In these time intervals the oscillation power reaches local minima.
The transition from the spiral to circular shape occurs by a decay of the spiral arms into several patches. Possibly because of the lower power of the oscillations, it is simply impossible to resolve the actual spiral shape in these instants of time, but there is a clear correlation between the clear quasi-spirality of the wave fronts and its high amplitude.
For example, during the power peaks (labels 2 and 3 in Fig.~\ref{fig2}a), the wave front shape is a spiral, while during the lower power (labels 1, 4 and 5 in Fig.~\ref{fig2}a), the fronts have a patchy circular shape. During the transition from the patchy circular shape to the spiral, the increase in the oscillation power is accompanied with the filling-up of the space between the patches, their coalescence, and transformation of the circular front into spiral arms. 

\subsection{Height structure of the wave fronts}

\begin{figure}[htpb]
   \centering
  \includegraphics[width=9.0 cm]{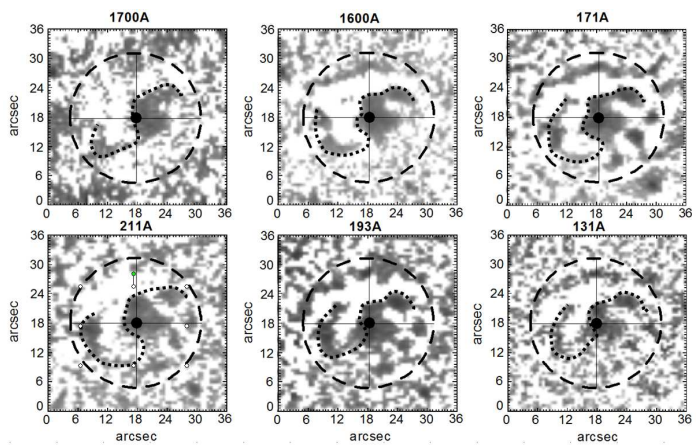}
  \caption{Wave fronts that indicate the spatial locations of 3-min oscillation power enhancements obtained {in different bands:  at 1700\AA\ and 1600\AA\ at 02:48:36~UT, at 171\AA\ and  211\AA\ at 02:49:00~UT, and at 193\AA\ and 131\AA\  at 02:49:12~UT}.
The signal is shown in the logarithmic scale. The dark dashed curve indicates the umbral boundary. The black dotted curves highlight the wave fronts. {The black indicates the oscillation power enhancement, white shows the lower power.}
  }
  \label{fig4} 
  \end{figure}

The horizontal structure of the 3-min wave-front quasi-spirality found in Sec.~\ref{2dan} in the 304\AA\ bandpass was studied in other bandpasses of AIA data. The data processing was performed by the PWF method, and was similar to what we described in Sec.~\ref{2dan}. Fig.~\ref{fig4} shows 3-min narrowband snapshots of the oscillation peer spatial distribution for different AIA bandpasses that measure the emission generated at different levels of the solar atmosphere. The spiral shape is evident at all the heights. The only exception is the 335\AA\ bandpass (not shown in Fig.~\ref{fig4}), which corresponds to the emission of a flaring high-temperature plasma. The lack of this effect in the  335\AA\ bandpass can be attributed to the low signal in that bandpass during the observations performed during the quiet-Sun time interval. The spiral is best seen in the 304\AA, 171\AA\ and 193\AA\ bandpasses, in which 3-min oscillations have the highest power. The snapshots obtained in different bandpasses are different from each other in
minor details, but the main two-armed spiral shape is same in all the images. In addition, all the images have the same circular wave front situated near the umbra-penumbra boundary. As the brightness of the signal decreases with the distance from the sunspot centre, the details seen in the penumbra should be attributed to enhancement of the noise by the logarithmic scale visualisation, and hence can be disregarded.

\subsection{Evolution of 3-min oscillations in time}
\label{ev3m}

The phase relationship between fronts of 3-min signals measured in different EUV channels was studied by calculating the 2D cross-correlation function for pairs of 3-min narrowband images of the wave amplitude spatial distribution. The function \textit{correl\_images.pro} of the AstroLib package\footnote{http://idlastro.gsfc.nasa.gov/} was used in the calculations. The image made in the instant of time that corresponded to the highest power of the oscillations at 304\AA, 02:49~UT (see Fig.~\ref{fig1}a) was taken as the base image. The 2D cross-correlation function was calculated for the base image and other images made for different instants of time and at different wavelengths. No shift in the spatial domain was applied. 

\begin{figure}
   \centering
  \includegraphics[width=9.0 cm]{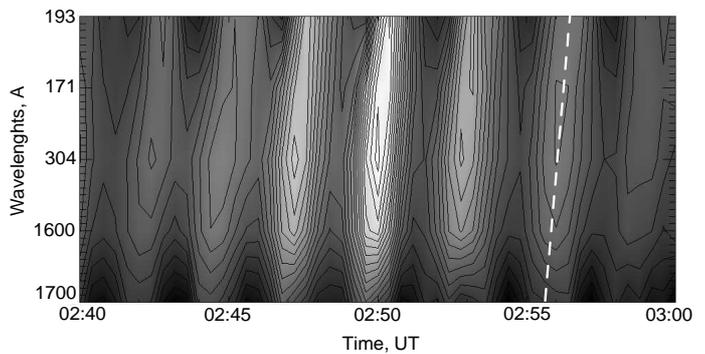}
  \caption{
2D cross-correlation function calculated for one train of 3-min oscillations, observed at different wavelengths. The base signal is the image taken at 02:49~UT at 304\AA\ when the spiral wave front was most pronounced. The value of the cross-correlation function is shown by the brightness. The time lag in the formation of the spiral wave front is highlighted by the dashed line. The wavelengths of different observational channels are shown in \AA.}
  \label{fig5}
  \end{figure}

Fig.~\ref{fig5} shows the 2D cross-correlation function calculated during one train of 3-min oscillations,
02:40--03:00~UT for different bandpasses. 
There is a systematic shift of the local maximum correlation (the brightness bump in Fig.~\ref{fig5}) in time with the observational bandpass (or the height). The quasi-spirality that corresponds to the local enhancement of the correlation first appears at the photospheric level (1700\AA) and then at higher levels: 1600\AA, 304\AA, 171\AA, and finally 193\AA. The time difference between the signals at 1700\AA\ and 193\AA\ is about 46 s. The phase displacement in time is highlighted in Fig.~\ref{fig5} by the dashed line. Thus, the spiral fronts begin at the photospheric level and gradually propagate upwards, reaching the corona. This finding is consistent with the single-pixel measurements made in \cite{2012ApJ...746..119R}. 

The time shift between the signals, corresponding to the local maxima of the 2D correlation function at different heights, allows for the estimating the vertical phase speed of the signal. Using the empirical model of an umbral atmosphere developed by \cite{1986ApJ...306..284M} and assuming that the 1700\AA\ emission comes from the height of 500~km above the photosphere, and 304\AA\ at about 2,200 km, we estimate the speed to be about 100 km/s. This value is slightly higher than the estimates of \cite{2012ApJ...746..119R} (about 70 km/s), \cite{2011SoPh..270..175A} (about 60 km/s) and \cite{2013A&A...554A.146K}  ($55\pm10$~ km/s). In these papers the speed was estimated for smaller RoI where the 3-min oscillation power was the highest, without accounting for the spatiotemporal evolution of the oscillations. In all cases the estimates are affected by the uncertainties in determining the specific height of the emission measured in a certain bandpass, strong variation of the sound speed with height, and also by the height extension of the coronal sources, and thus have only an illustrative character. 

\begin{figure}
   \centering
  \includegraphics[width=9.0 cm]{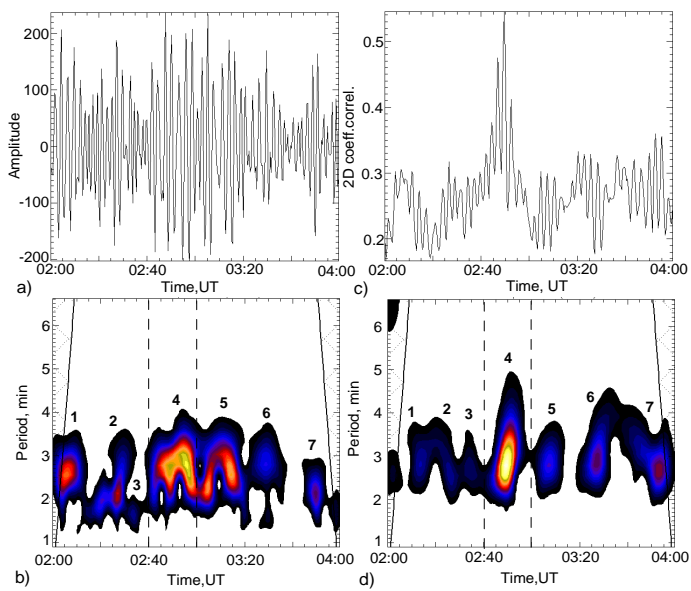}
  \caption{ Time evolution of the 3-min oscillation amplitude (a), and the 2D cross-correlation coefficients (c) during two hours (02:00--04:00~UT) at 304\AA\ .  The distribution of the spectral power of these quantities in the narrowband is shown in panels (b) and (d), respectively. Time instants of the enhanced oscillation power are enumerated.
  }
  \label{fig6}
  \end{figure}

For a longer time interval, 02:00--04:00~UT, the relationship between the variation of the 3-min oscillation power and the 2D cross-correlation of the wave fronts is demonstrated by Fig.~\ref{fig6}. The figure shows the 3-min signal and the time variation of the 2D cross-correlation function calculated for the base image and the images taken at 1600\AA\ and their wavelet power spectra. The 3-min signal is taken by summation of the 304\AA\ signals at five neighbouring pixels at the umbra centre.
The wave train nature of 3-min oscillations, with the 12--20~min period, is seen during the whole observation. The wavelet spectrum shows some frequency drift within the 2--4~min period range in the individual trains, which is consistent with the earlier findings \citep{2012A&A...539A..23S}. The wavelet spectra correlate well for the amplitude of 3-min oscillation trains and the 2D cross-correlation function. In particular, increase in the amplitude is accompanied with the increase in the  2D cross-correlation function. Thus, the effect of the increase in the signal quasi-spirality with the increase in the oscillation amplitude is persistent.

\subsection{Evolution of wave fronts during one cycle of 3-min oscillations}
\label{evcyl}

As pointed out in Sec.~\ref{ev3m}, the quasi-spirality of the 3-min wave fronts evolves with a 3-min periodicity.
We analysed this process for one oscillation cycle, 02:49:10--02:52:10~UT, when the oscillations reached the highest power and the wave front quasi-spirality was well developed.

\begin{figure}[htpb]
   \centering
  \includegraphics[width=9.0 cm]{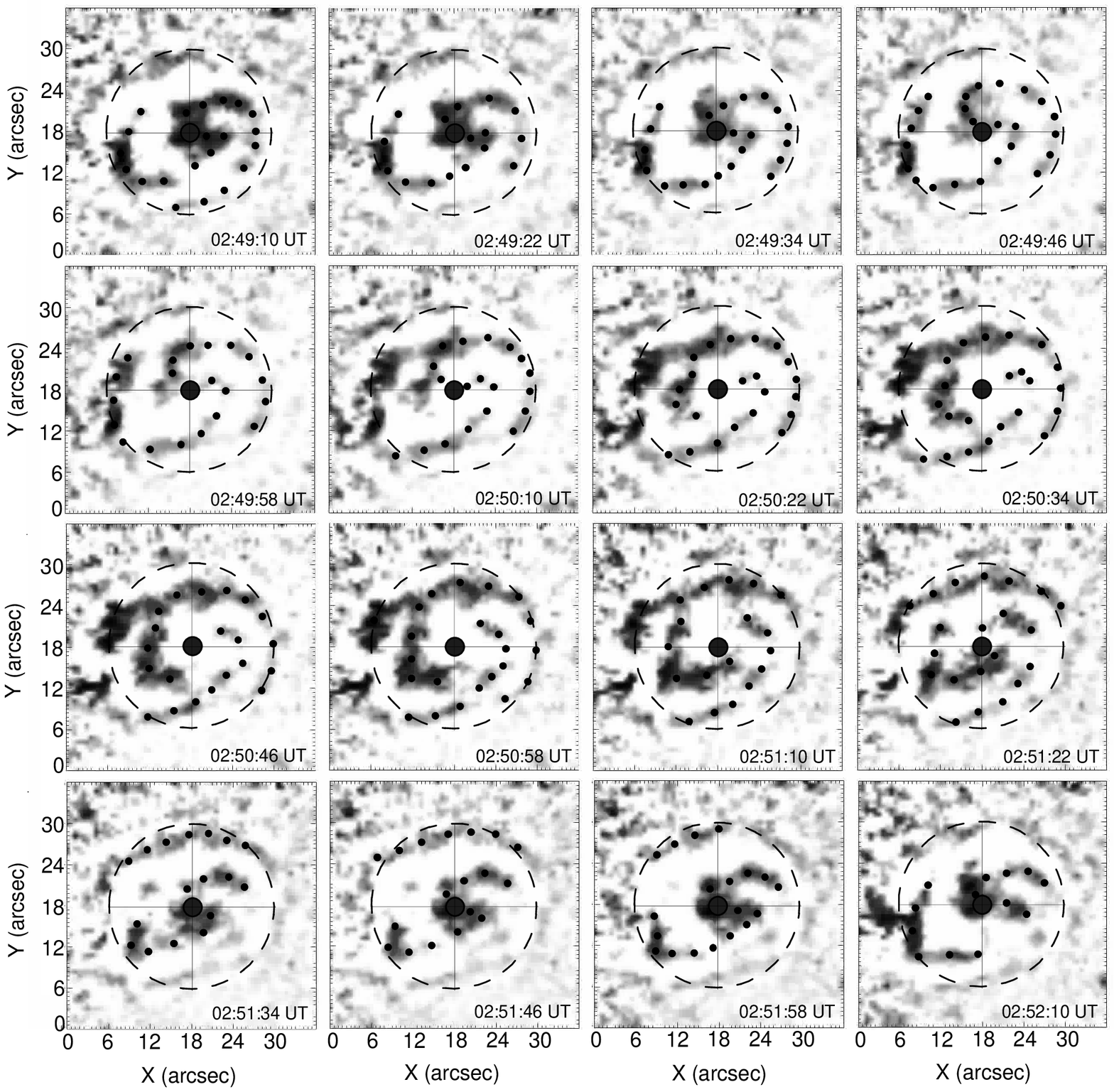}
  \caption{Different phases of the apparent horizontal wave fronts during one cycle of 3-min oscillations at 304\AA.
  The black dots highlight the wave fronts. The dashed line shows the umbral boundary.  
  }
  \label{fig7}
  \end{figure}

Fig.~\ref{fig7} shows the snapshots of the wave front evolution in time from one maximum to the next, obtained at 304\AA. It is evident that the first maximum of the oscillation, at 02:49:10~UT, is accompanied by the appearance of well-developed two-armed spiral originated at the umbra centre (c.f. Figs.~\ref{fig3}a and \ref{fig4}). During the decrease in the signal amplitude the arms experience anti-clockwise rotation and radial movement. Simultaneously, we see the development of a void spiral that highlights the signal minima. The void spiral becomes fully developed at 02:50:34--02:50:46~UT. Its shape resembles the shape of the bright spiral that was made of the wave fronts constructed by the signal maxima. Then the process repeats and a clearly visible bright spiral is formed by 02:51:58~UT. 

Similar dynamics is evident in the 1700\AA, 1600\AA, 171\AA, 193\AA, 211\AA, and 131\AA\ bandpasses. There are some minor discrepancies. For example, at the temperature minimum level (1700\AA), we see the central part of the spiral wave fronts, while the circular fronts appearing after the break-down of the spiral in other bands are hardly visible. The spiral shape is most pronounced at higher levels, at 304\AA, 171\AA, 193\AA, and 211\AA. The geometrical horizontal size of the 3-min oscillations region increases with height, similarly to the findings in  
\cite{2012ApJ...756...35R} and \cite{2014A&A...561A..19Y}. In the corona (e.g. at 171\AA) 3-min oscillations show an additional geometrical feature: they highlight the field-aligned plasma channels that guide the oscillations in the corona. The channels are extended radially from the umbra over the penumbra and farther out.

\subsection{Spectral structure of the 3-min oscillation source}

The power spectrum of umbral oscillations, shown in Fig.~\ref{fig2}b, demonstrates that the 3-min oscillation peak is broad and extends approximately from 4~mHz to 11~mHz (from 4~min to 1.5~min). The spectral power enhancement in this band varies and consists of several distinct narrow peaks. In particular, together with the main peak at 2.8~min, there are at least two other peaks, at 1.9~min and 2.3~min. The question is whether the evolution of the 3-min wave fronts, discussed in Sec.~\ref{evcyl}, is associated with some fine changes in the signal spectrum, and whether different spatial details, such as the spiral arms and the central patch, are associated with different peaks in the 3-min spectral power enhancement.
An additional motivation of this study is the dependence of the 3-min oscillation sources on the acoustic cut-off frequency variation in the horizontal direction, which can be different across the umbra \citep[see e.g.][for a discussion]{2008SoPh..248..395S, 2014A&A...561A..19Y}. 

\begin{figure}
   \centering
  \includegraphics[width=9.0 cm]{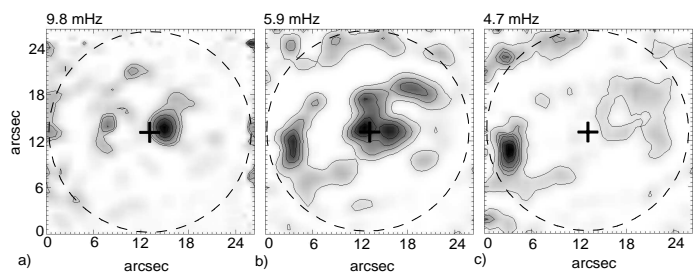}
  \caption{Spatial distribution of the narrowband power of 3-min oscillations at 304\AA\ in the sunspot umbra at 02:49:10~UT: (a) 9.8~mHz (1.7~min), (b) 5.9~mHz (2.8~min), and (c) 4.7~mHz (3.5~min).
  The dashed line shows the umbral boundary. The sunspot centre is indicated by the cross. 
  }
\label{fig8}
  \end{figure}

The fine spectral structure of the spatial distribution of 3-min umbral oscillations was studied the use of the PWF-analysis.
Narrowband amplitude maps of the oscillations recorded at 304\AA\ were constructed in the spectral range from 1.0~min to 4.0~min with the spectral resolution of 0.1~min, at the time of the highest oscillation power, 02:49~UT. The results obtained revealed the dependence of the wave front geometry on the oscillation period (Fig.~\ref{fig8}). 

We consider the 3-min wave source evolution with the period increase.
For 1.5~min period the oscillation is localised near the umbra centre (Fig.~\ref{fig8}a). The oscillation source is almost circular, of 4\arcsec\  in diameter. For longer periods its diameter grows, and the source becomes stronger (or brighter). The maximum is reached at a period of 1.8~min, which corresponds to the short-period peak in the 3-min spectrum. Simultaneously, two arc-shaped details develop and become most pronounced at 2.6~min. For shorter periods these details are disconnected from the central patch. For longer periods, the inner ends of the arcs reach the central patch and merge it. At a period of 2.8~min, corresponding to the spectral maximum of the 3-min oscillations, we see two well-developed spiral wave fronts linked with the central patch (Fig.~\ref{fig8}b). At longer periods, the strength of the central patch rapidly decreases, and the arms of the spirals become disconnected from  each other. Thus, there remain two disconnected arc-shaped fronts (Fig.~\ref{fig8}c), which for longer periods decline as well. Then, at the periods close to the 5-min peak in the spectrum, the strength of the arc-shaped fronts again increases, and they form narrow annuli at the umbra-penumbra boundary. 

Thus, the narrowband analysis shows that the observed quasi-spirality of 3-min oscillation sources results from the superposition of two narrowband sources, the short-period (about 1.8~min) circular patch at the centre of the umbra, and two arcs or a two-armed spiral at the  2.5--3.1~min periods. The same analysis made for other time intervals gives the same spatio-spectral structure. 

\begin{figure}
   \centering
  \includegraphics[width=9.0 cm]{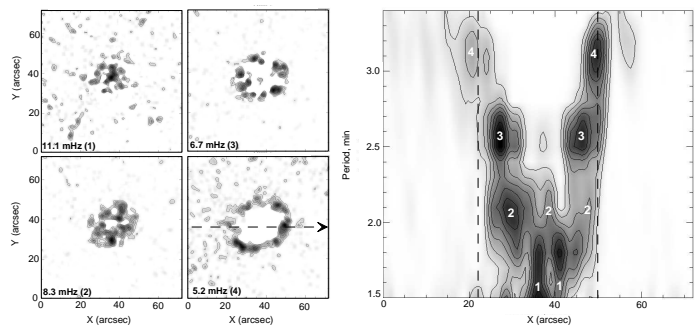}
  \caption{(a) Running-difference narrowband maps of the umbral oscillations signals obtained at 304\AA, integrated from 02:30 to 03:30~UT, at the frequencies 11.11~mHz (1.5~min), 8.3~mHz (2.0 min),  6.7~mHz (2.5~mHz), and 5.2~mHz (3.2~min). (b) Dependence of the narrowband power distribution across the umbra (in the horizontal direction via the sunspot centre, see the dashed arrow in the 5.2~mHz map in panel a) on the oscillation period. The vertical dashed lines show the umbral boundaries. The digits indicate the spectral power enhancements that coincide with the spectral peaks in Fig.~\ref{fig2}b.
  }
  \label{fig9}
  \end{figure}

Above we discussed the instant spatio-spectral structure of umbral oscillations. Consider the spatio-spectral structure of 3-min oscillations at 304\AA, averaged over a long period of time, for example 02:30--03:30~UT, aiming to derive the average location of the different narrowband spatial details in the umbra. The spectral band from 1.5~min to 3.4~min was split into spectral steps of 0.1~min. The slow variations of the details were emphasised by the use of the running image difference signals. The results are shown in Fig.~\ref{fig9}, which shows it is evident that high-frequency oscillations, with period of about 1.5~ min, are localised at the umbra centre. With the decrease in the frequency, the oscillations are localised in the annuli of the increasing radius around the umbra centre. At the low-frequency end of the 3-min band, the radius of the annulus is largest. Fig.~\ref{fig9}b demonstrates that the annulus radii change smoothly with the oscillation period. The blobs in the dependence of the annulus radius on the period correspond to the spectral peaks in the 3-min band, see Fig.~\ref{fig2}b. These values of the periods were chosen for the individual panels of Fig.~\ref{fig9}a.

\section{Results}
\label{res}

We analysed fine horizontal, vertical, temporal, and spectral structure of umbral oscillations in a circular well-developed sunspot situated in the Northern hemisphere near the central meridian that was observed with SDO/AIA. The analysis was performed by time-distance mapping and the PWF technique.
Below we summarise our results.

The power of 3-min oscillations integrated over the umbra evolves in time, forming wave trains of 12--20 min duration. The spectrum of the 3-min oscillations has a fine structure. 
Time-distance maps constructed along slits passing through the sunspot centre show that in the time intervals of the enhanced oscillation power, 3-min wave fronts are mainly parallel to each other. In the time intervals of lower oscillation power, the wave fronts become irregular: they experience breaks, dislocations and phase shifts. 

A 2D analysis of narrowband umbral oscillations observed at 304\AA\ revealed spatial structuring of the oscillations. Wave fronts have the form of an evolving two-armed spiral and a stationary circular patch at the spiral origin, situated near the umbra centre. In time, the spiral rotates anti-clockwise. The width of the arms increases when they approach the umbra-penumbra boundary. When the oscillation power decreases, the arms become faint, decompose to a series of individual patches, and their shape become circular. This evolution scenario is consistent with the behaviour revealed by the time-distance map.
The 2D analysis of narrowband umbral oscillations performed at other wavelengths showed that the quasi-spirality of the wave front occurs at all heights, from the temperature minimum (1700\AA) to the corona (193\AA). 
The quasi-spirality is not seen in high-temperature bandpass, 335\AA. The 2D cross-correlation function constructed for the signals at different bandpasses shows that there is a time lag between the signals at the lower and higher levels of the solar atmosphere. This finding indicates that the oscillations are associated with upward propagating waves. The phase speed estimated by the time lag between the signal at 304\AA\ and 171\AA\ is about 100~km/s. This value is consistent with the previously obtained estimations, which account for the uncertainty in the height of the 171\AA\ emission.

The fine time evolution of the horizontal structure of 3-min wave fronts at 304\AA\ shows that the spiral structure experiences periodic changes during one cycle of 3-min oscillations. In the high-power phases of the oscillation, the quasi-spirality is most pronounced. In low power phases, the quasi-spirality is better seen in the signal absence, the void of the signal. 

The analysis of the fine spectral structure of the horizontal distribution of the umbral oscillation power performed at 304\AA revealed that the quasi-spirality results from the combination of spatially separated details that correspond to oscillations with different periods. The central circular patch is associated with higher frequency oscillations with a period of about 1.8~min. The spiral arms are arc-shaped patches of enhanced 2.6--3.1-min oscillations.

Averaging the narrowband signals over a one-hour time interval showed that their spatial structure is concentric annuli whose centres coincide with the umbra centre. Annuli with the smallest diameters are obtained for the highest frequencies. The annulus diameter increases with the decrease in frequency. The annulus diameter changes continuously with the frequency. The signals in some annuli are stronger than in the other, which corresponds to the fine spectral structure of the 3-min enhancement in the integrated spectrum of umbral oscillations. 


\section{Discussion}
\label{disc}	
	
The observed global horizontal structure of umbral oscillations is consistent with an interpretation in terms of the model based on the acoustic cut-off frequency. According to this model \citep[see, e.g.][]{2008SoPh..251..501Z, 2011ApJ...728...84B, 2012A&A...539A..23S}, a broadband initial perturbation develops in a quasi-periodic wave train with the local acoustic cut-off frequency $f_\mathrm{ac} \propto \cos \theta / \sqrt{T}$ in the strong magnetic field of the umbra, where $\theta$ is the angle between the magnetic field and the vertical and $T$ is the plasma temperature, see \citet{1977A&A....55..239B} for the basic theory, and \citet{2012A&A...537A..96K} for example, and references therein for recent numerical simulations. Thus, assuming that the magnetic field lines in the sunspot atmosphere expand horizontally from the sunspot axis, the frequency $f_\mathrm{ac}$ decreases in the horizontal direction with the distance from the sunspot axis. This naturally explains that the umbral oscillations of low-frequencies occupy the annuli of increasing radius and width \citep[see also discussion in][]{2014A&A...561A..19Y}. At the umbra centre the magnetic field lines are almost vertical. This explains why the oscillations have the highest frequency in the central patch. It is not clear why the radial dependence on the oscillation power varies, in contrast with the smooth radial decrease in the frequency. As the highest power annuli correspond to the peaks in the 3-min band, this question reduces to why umbral oscillations have a fine spectral structure. 

The obtained vertical structure of the oscillations is consistent with the filter model. The growing time lag between the oscillations seen at subsequently higher levels of the atmosphere demonstrates the propagating nature of the waves, which results in the generation of propagating longitudinal waves in coronal sunspot fans. The established good correlation of the wave front shapes at different heights clearly indicates the collective character of the oscillations
in the vertical direction. 

Non-uniformity of the wave fronts in the polar angle, around the sunspot vertical axis, is not explained by the available theory. 
The question remains whether the observed spirality of the spatial distribution of the 3-min oscillation power in the umbra is associated with the twist of the magnetic field in the sunspot. {The study of the magnetic geometry over this sunspot by \cite{2012ApJ...756...35R} did not show that the field was twisted. Moreover, because in the low-beta plasma of sunspots the magnetic twist should be uniform in the azimuthal direction because of  the force balance, it cannot contribute to the wave front or oscillation power non-uniformity in the azimuthal direction.  Thus, we disregard the association of the observed phenomenon with the magnetic twist.}
The individual arcs and spiral arms can be attributed to the local dependence of the magnetic field inclination on the polar angle. For example, the observed behaviour can occur if the field lines anchored at the same radius form the umbra centre, but at different polar angles, have different inclination (see, e.g. Fig.~5 of \cite{2012ApJ...756...35R}). In that case, the annulus shapes of the wave fronts can be deformed into arcs and even segments of a spiral arm in some polar angle sector. However, this scenario cannot explain why the detected spirals have two arms of the same rotation sense, or similar arcs at the opposite polar angles. Explanation of the fine polar-angle structure of umbral oscillations constitutes an interesting theoretical problem.

\begin{acknowledgements}
This work is supported by the STFC Warwick Astrophysics Consolidated Grant ST/L000733/1, the European Research Council under the \textit{SeismoSun} Research Project No. 321141, the BK21 plus program through the National Research Foundation funded by the Ministry of Education of Korea (VMN), the Marie Curie PIRSES-GA-2011-295272 \textit{RadioSun} project, and the Russian Foundation of Basic Research under grant 13-02-00044 (RS, VMN) and grants13-02-90472, 13-02-1000 and 14-0291157 (RS).
\end{acknowledgements}

\bibliographystyle{aa} 

\end{document}